\documentclass{article}
\usepackage{amsmath}

\input{tcilatex}
\begin{document}

\centerline{ \large\bf Reply to the comment on "Stochastic local
operations and} 
\centerline{\large\bf  classical communication
invariant and } 
\centerline{ \large\bf the residual entanglement for
n qubits"}

\footnote{%
The paper was supported by NSFC(Grants No. 60433050 and 60673034), and the
basic research fund of Tsinghua university NO: JC2003043.}

\centerline{Dafa Li$^{a}$\footnote{email
address:dli@math.tsinghua.edu.cn},  Xiangrong
Li$^{b}$, Hongtao Huang$^{c}$, Xinxin Li$^{d}$ }

\centerline{$^a$ Department of mathematical sciences, Tsinghua
University, Beijing 100084 CHINA}

\centerline{$^b$ Department of Mathematics, University of
California, Irvine, CA 92697-3875, USA}

\centerline{$^c$ Electrical Engineering and Computer Science Department} %
\centerline{ University of Michigan, Ann Arbor, MI 48109, USA}

\centerline{$^d$ Department of computer science, Wayne State
University, Detroit, MI 48202, USA}


Coffman et al. presented the residual entanglement for three qubits \cite%
{Coffman}.\ In \cite{LDF07}, we proposed the residual entanglement for any
state $|\psi \rangle =\sum_{i=0}^{2^{n}-1}a_{i}|i\rangle $ for odd $n$
qubits. For readability, we repeat our definition on page 4 of \cite{LDF07}\
as follows.

When $n$\ is odd, by means of the invariant for odd $n$ qubits, we define
that for any state $|\psi \rangle $, the residual entanglement\textbf{\ }%
\begin{equation}
\tau (\psi )=4|(\overline{IV}(a,n))^{2}-4IV^{\ast
}(a,n-1)IV_{+2^{n-1}}^{\ast }(a,n-1)|.  \label{oddre1}
\end{equation}%
\ 

Then, we derived corollary 2 in \cite{LDF07}. That is, if $|\psi \rangle $
and $|\psi ^{\prime }\rangle $ are equivalent under SLOCC, then 
\begin{equation}
\tau (\psi )=\tau (\psi ^{\prime })|\det^{2}(\alpha )\det^{2}(\beta
)\det^{2}(\gamma )...|.  \label{corollary}
\end{equation}

As well known, the residual entanglement for three qubits or 3-tangle is
invariant under permutations of the three qubits \cite{Coffman}. Then, did\
they want to indicate that our residual entanglement is not invariant under
permutations of all the odd $n$ qubits? For this purpose, logically it only
needs a counter-example. Whereas, we proved in \cite{LDF07b} that our
residual entanglement $\tau $ for odd $n$ qubits has the following
properties: (1). $0\leq \tau \leq 1$. (2). $\tau $ is invariant under
SL-operators, especially LU-operators. (3). $\tau $ is an entanglement
monotone. (4). $\tau $ is invariant under permutations of qubits $2$, $3$,
..., $n$. (5). for product states, $\tau =0$ or is multiplicative. To show
that our residual entanglement $\tau $ for odd $n$ qubits is not invariant
under permutations of all the odd $n$ qubits, we gave a simple example in 
\cite{LDF07b}. See example 5 in \cite{LDF07b}. The example is $|\psi \rangle
=(1/2)(|0\rangle +|7\rangle +|24\rangle +|31\rangle )$. By our definition
for five qubits, a simple calculation shows that $\tau (\psi )=0$. Under the
permutation of the qubits (1 $\leftrightarrow $ 5), $|\psi \rangle $ becomes 
$|\psi ^{\prime }\rangle =(1/2)(|0\rangle +|9\rangle +|22\rangle +|31\rangle
)$. However, $\tau (\psi ^{\prime })=1$. This is a weakness of our proposal.

In this reply, we improve our definition in Eq. (\ref{oddre1})\ to overcome
this weakness as follows. Let us recall that $\tau (\psi )$ is invariant
under permutations of qubits $2$, $3$, ..., $n$ \cite{LDF07b}. Then, what
will happen to $\tau (\psi )$\ under the transposition $(1,i)$ of qubits $1$
and $i$?\ Let $|\psi ^{\prime }\rangle $ be obtained from $|\psi \rangle $
under the permutation $\sigma $\ of the qubits, and let us write $|\psi
^{\prime }\rangle =\sigma |\psi \rangle $. By means of $\tau (\psi )$ in Eq.
(\ref{oddre1}), let

\begin{equation}
\tau ^{(i)}(\psi )=\tau ((1,i)\psi ),i=2,3,...,n,  \label{def1}
\end{equation}
and $\tau ^{(1)}(\psi )=\tau (\psi )$. Then, $\tau ^{(i)}(\psi )$, $i=1$, $2$%
, ... , $n$, are invariant under any permutation of the qubits: $1$, $2$,
..., $(i-1)$, $(i+1)$, ..., $n$ by the following property 1, and $\tau
^{(i)}(\psi )$ satisfy corollary 2 in Eq. (\ref{corollary}) by the following
property 4.

Let $R(\psi )=\frac{1}{n}\sum_{i=1}^{n}\tau ^{(i)}(\psi )$. Then, $R(\psi )$
is invariant under any permutation of all the odd $n$\ qubits by the
following property 3, and $R(\psi )$ satisfies corollary 2 in Eq. (\ref%
{corollary}) by the following property 4. It can be verified that $R(\psi )$
also satisfies: (1). $0\leq R\leq 1$; (2). $R$ is invariant under
SL-operators, especially LU-operators; (3). $R$ is an entanglement monotone.
However, for some product states, $R(\psi )$ is not multiplicative. For
example, let $|\psi \rangle =(1/2)((|00\rangle +|11\rangle )_{12}\otimes
(|000\rangle +|111\rangle )_{345})$. Then, $R(\psi )=3/5$.

Now, let us redefine the residual entanglement for odd $n$ qubits or the odd 
$n$-tangle instead of $\tau (\psi )$ as follows. When $n$\ is odd, by means
of $\tau (\psi )$,\ we define that for any state $|\psi \rangle $, the
residual entanglement

\begin{equation}
R(\psi )=\frac{1}{n}\sum_{i=1}^{n}\tau ^{(i)}(\psi ).  \label{new-def}
\end{equation}

Next let us see the performance of $R(\psi )$ for three qubits and five
qubits.

Let $n=3$ in Eq. (\ref{oddre1}) above. Then the definition for three qubits
in \cite{LDF07} is repeated as follows.

\begin{equation}
\tau (\psi
)=4|((a_{0}a_{7}-a_{1}a_{6})-(a_{2}a_{5}-a_{3}a_{4}))^{2}-4(a_{0}a_{3}-a_{1}a_{2})(a_{4}a_{7}-a_{5}a_{6})|.
\label{res-3-qubits}
\end{equation}%
\ In Remark 1 of \cite{LDF06}, we indicated that our Eq. (\ref{res-3-qubits}%
) happens to be Coffman et al.'s residual entanglement for three qubits,
which is $\tau _{ABC}=4\left\vert d_{1}-2d_{2}+4d_{3}\right\vert $, where
the expressions for $d_{i}$ are omitted here. This fact can be verified\ by
expanding Eq. (\ref{res-3-qubits}) suggested by the reviewer of \cite{LDF06}%
. We also showed (see (5) of p. 429, \cite{LDF06}) 
\begin{eqnarray}
&&((a_{0}a_{7}-a_{1}a_{6})-(a_{2}a_{5}-a_{3}a_{4}))^{2}-4(a_{0}a_{3}-a_{1}a_{2})(a_{4}a_{7}-a_{5}a_{6})=
\label{def-3-qubits-1} \\
&&((a_{0}a_{7}-a_{3}a_{4})+(a_{1}a_{6}-a_{2}a_{5}))^{2}-4(a_{3}a_{5}-a_{1}a_{7})(a_{2}a_{4}-a_{0}a_{6})=
\label{def-3-qubits-2} \\
&&(a_{0}a_{7}-a_{3}a_{4}-(a_{1}a_{6}-a_{2}a_{5}))^{2}-4(a_{1}a_{4}-a_{0}a_{5})(a_{3}a_{6}-a_{2}a_{7})
\label{def-3-qubits-3}
\end{eqnarray}

When $n=3$, $\tau ^{(1)}(\psi )=\tau (\psi )$; under the transposition $%
(1,2) $ of qubits $1$ and $2$, $\tau (\psi )$ becomes $4|$Eq. (\ref%
{def-3-qubits-3})$|$, i.e., $\tau ^{(2)}(\psi )=4|$Eq. (\ref{def-3-qubits-3})%
$|$; under the transposition $(1,3)$ of qubits $1$ and $3$, $\tau (\psi )$
becomes $4|$Eq. (\ref{def-3-qubits-2})$|$, i.e., $\tau ^{(3)}(\psi )=4|$Eq. (%
\ref{def-3-qubits-2})$|$. By Eqs. (\ref{def-3-qubits-1}), (\ref%
{def-3-qubits-2}), and (\ref{def-3-qubits-3}), $\tau (\psi )=\tau
^{(1)}(\psi )=\tau ^{(2)}(\psi )=\tau ^{(3)}(\psi )$. Thus, $R(\psi )=$ $%
\tau (\psi )$. That is, $R(\psi )$ is just Coffman et al.'s residual
entanglements for three qubits or 3-tangle.\ 

When $n=5$, they said that their $Z_{12345}^{1}$ is our $\tau (\psi )$ for
five qubits. It can be verified that under the transpositions $(1,i)$ of
qubits $1$ and $i$, where $i=2,3,4,5$, our $\tau (\psi )$ for five qubits
becomes their $Z_{12345}^{2}$, $Z_{12345}^{3}$, $Z_{12345}^{4}$, $%
Z_{12345}^{5}$, respectively, and under the transposition $(i,j)$ of qubits $%
i$ and $j$, $Z_{12345}^{i}$ becomes $Z_{12345}^{j}$ and vice versa. Thus, $%
R(\psi )=\frac{1}{5}\sum_{i=1}^{5}Z_{12345}^{i}$. That is, $R(\psi )$ is an
average of their $Z_{12345}^{i}$, $i=1,2,3,4,5$.\ 

However, their argument does not seem to discuss the above weakness.\ They
said that their $Z_{12345}^{i}$, $i=1,2,3,4,5$, also satisfy our corollary 2
in Eq. (\ref{corollary}), and thought that this is the weakness of our
proposal. Now, instead of $\tau (\psi )$\ we define $R(\psi )$ in Eq. (\ref%
{new-def}) as the residual entanglement for odd $n$ qubits. $R(\psi )$
represents a collective property of all the odd $n$ qubits, while their $%
Z_{12345}^{i}$, $i=1,2,3,4,5$, are not invariant under permutations of all
the odd $n$ qubits by property 1. Note that $a_{13}a_{28}$ in their $%
Z_{12345}^{i}$, $i=1,2,3,4,5$, should be $a_{13}a_{18}$, $a_{5}a_{15}$ in
their $Z_{12345}^{1}$ should be $a_{5}a_{10\text{ }}$.

Property 1.

$\tau ^{(i)}(\psi )$, $i=1,2,...,n$, are invariant under any permutation of
the qubits: $1$, $2$, ..., $(i-1)$, $(i+1)$, ..., $n$.

Proof. It is true for $\tau ^{(1)}(\psi )$ because $\tau ^{(1)}(\psi )=\tau
(\psi )$ and in \cite{LDF07b} we proved that $\tau (\psi )$ is invariant
under any permutation of the qubits: $2$, $3$, ..., $n$. Here, we show that $%
\tau ^{(2)}(\psi )$ also has this property. \ To show that $\tau ^{(i)}(\psi
)$ has the property, we only need to replace $(1,2)$ by $(1,i)$\ in the
proof for $\tau ^{(2)}(\psi )$. For $\tau ^{(2)}(\psi )$, let $\sigma $ be
any permutation of the qubits: $1$, $3$, $4$, ..., $n$. There are two cases.

Case 1. $\sigma (1)=1$. Thus, $\sigma $ can be considered as a permutation
of the qubits: $3$, $4$, ..., $n$. In this case, $\sigma =(1,2)(1,2)\sigma
=(1,2)\sigma (1,2)$ because $\sigma $ and $(1,2)$ are disjoint. We argue
that $\tau ^{(2)}(\psi ^{\prime })=\tau ^{(2)}(\psi )$ as follows.

$\tau ^{(2)}(\sigma |\psi \rangle )=\tau ^{(2)}((1,2)\sigma (1,2)\psi )$

$\ \ \ \ \ =\tau ((1,2)(1,2)\sigma (1,2)\psi )$ by Eq. (\ref{def1})

$\ \ \ \ \ =\tau (\sigma (1,2)\psi )$

$\ \ \ \ =\tau ((1,2)\psi )$ by the property of $\tau (\psi )$

\ \ \ \ \ \ $=\tau ^{(2)}(\psi )$ by Eq. (\ref{def1}).

Case 2. $\sigma (1)\neq 1$. It is known that every permutation is a product
of disjoint cycles. Let $\sigma
=(1,i_{1},i_{2},...,i_{l_{1}})(j_{1},...,j_{l_{2}})...(t_{1},...,t_{l_{s}})$%
, where these cycles are disjoint. Clearly, $%
(1,i_{1},i_{2},...,i_{l_{1}})=(1,2)(2,i_{1},i_{2},...,i_{l_{1}})(1,2)$.

$\tau ^{(2)}(\psi ^{\prime })=$ $\tau
^{(2)}((1,2)(2,i_{1},i_{2},...,i_{l_{1}})(1,2)(j_{1},...,j_{l_{2}})...(t_{1},...,t_{l_{s}})\psi ) 
$

\ \ \ \ \ \ \ $=\tau
((2,i_{1},i_{2},...,i_{l_{1}})(1,2)(j_{1},...,j_{l_{2}})...(t_{1},...,t_{l_{s}})\psi ) 
$ by Eq. (\ref{def1})

\ \ \ \ \ \ \ $=\tau ((1,2)(j_{1},...,j_{l_{2}})...(t_{1},...,t_{l_{s}})\psi
)$ by the property of $\tau (\psi )$

\ \ \ \ \ \ \ $=\tau
^{(2)}((j_{1},...,j_{l_{2}})...(t_{1},...,t_{l_{s}})\psi )$ by Eq. (\ref%
{def1})

\ \ \ \ \ \ \ $=\tau ^{(2)}(\psi )$ by case 1.

Property 2. Let $(i,j)$ be a transposition. Then $\tau ^{(i)}((i,j)\psi
)=\tau ^{(j)}(\psi )$ and $\tau ^{(j)}((i,j)\psi )=\tau ^{(i)}(\psi )$. That
is, under the transposition $(i,j)$ of qubits $i$ and $j$, $\tau ^{(i)}(\psi
)$ becomes $\tau ^{(j)}(\psi )$ and vice versa.

Proof. First let us prove that $\tau ^{(i)}((i,j)\psi )=\tau ^{(j)}(\psi )$.
By Eq. (\ref{def1}), $\tau ^{(i)}((i,j)\psi )=\tau ((1,i)(i,j)\psi )$ $=\tau
(1,i,j)\psi )$. By property 1, $\tau ^{(j)}(\psi )=\tau ^{(j)}((1,i)\psi )=$ 
$\tau ((1,j)(1,i)\psi )$ (by Eq. (\ref{def1})) $=\tau (1,i,j)\psi )$. Hence, 
$\tau ^{(i)}((i,j)\psi )=\tau ^{(j)}(\psi )$.

Next, let us prove that $\tau ^{(j)}((i,j)\psi )=\tau ^{(i)}(\psi )$. By Eq.
(\ref{def1}), $\tau ^{(j)}((i,j)\psi )=\tau ((1,j)(i,j)\psi )=\tau
(1,j,i)\psi )$. By property 1, $\tau ^{(i)}(\psi )=\tau ^{(i)}((1,j)\psi )=$ 
$\tau ((1,i)(1,j)\psi )$ (by Eq. (\ref{def1})) $=\tau (1,j,i)\psi )$. Hence, 
$\tau ^{(j)}((i,j)\psi )=\tau ^{(i)}(\psi )$.

Property 3. $R(\psi )$ is invariant under any permutation of all the odd $n$
qubits.

Proof. Case 1. Let $(i,j)$ be a transposition. Let us prove that $%
R((i,j)\psi )=R(\psi )$ as follows. Note that $R((i,j)\psi )=\frac{1}{n}%
\sum_{i=1}^{n}\tau ^{(i)}((i,j)\psi )$. By property 2, $\tau
^{(i)}((i,j)\psi )=\tau ^{(j)}(\psi )$ and $\tau ^{(j)}((i,j)\psi )=\tau
^{(i)}(\psi )$. By property 1, $\tau ^{(k)}((i,j)\psi )=\tau ^{(k)}(\psi )$
when $k\neq i$ or $j$. Therefore, $R((i,j)\psi )=R(\psi )$.

Case 2. Let $\sigma $ be any permutation of all the odd $n$ qubits. Let us
prove that $R(\sigma \psi )=R(\psi )$ as follows. As well known, any
permutation is a product of transpositions. Thus, we can write $\sigma
=(i_{1},j_{1})(i_{2},j_{2})...(i_{s},j_{s})$. By case 1 and induction, it is
straightforward that $R((i_{1},j_{1})(i_{2},j_{2})...(i_{s},j_{s})\psi
)=R(\psi )$.

Property 4. If the states $|\psi ^{\prime }\rangle $ and $|\psi \rangle $
are equivalent under SLOCC, i.e., $|\psi ^{\prime }\rangle =A_{1}\otimes
...A_{i}\otimes \cdots |\psi \rangle $, then $\tau ^{(i)}(\psi )$ and $%
R(\psi )$\ satisfy corollary 2 in Eq. (\ref{corollary}).

Proof. When $i=1$, this is trivial because $\tau ^{(1)}(\psi )=\tau (\psi )$%
. Let $(1,i)$ be a transposition of qubits $1$ and $i$. Then, $(1,i)|\psi
^{\prime }\rangle =(1,i)A_{1}\otimes ...A_{i}\otimes \cdots |\psi \rangle
=A_{i}\otimes ...A_{1}\otimes \cdots (1,i)|\psi \rangle $. It means that $%
(1,i)|\psi ^{\prime }\rangle $ and $(1,i)|\psi \rangle $ are equivalent
under SLOCC. By corollary 2 in Eq. (\ref{corollary}), $\tau ((1,i)\psi
^{\prime })=\tau ((1,i)\psi )|\det^{2}(A_{1})\det^{2}(A_{2})...|$. Then, by
Eq. (\ref{def1}), $\tau ^{(i)}(\psi ^{\prime })=\tau ^{(i)}(\psi
)|\det^{2}(A_{1})\det^{2}(A_{2})...|$. It is easy to see that $R(\psi
^{\prime })=R(\psi )|\det^{2}(A_{1})\det^{2}(A_{2})...|$.

\end{document}